\documentclass[11pt,a4paper,twoside]{article}
\usepackage{amsmath}
\usepackage{amsfonts}
\usepackage{changes}
\usepackage{color}
\usepackage{graphicx}
\usepackage{lmodern}
\usepackage[T1]{fontenc}
\usepackage[utf8]{inputenc} 
\numberwithin{equation}{section}
\usepackage{comment}
\usepackage{tabularx}

\makeatletter
\newcounter {subsubsubsection}[subsubsection]
\renewcommand\thesubsubsubsection{\thesubsubsection .\@arabic
	\c@subsubsubsection}
\newcommand\subsubsubsection{\@startsection{subsubsubsection}{4}{\z@}%
	{-3.25ex\@plus -1ex \@minus -.2ex}%
	{1.5ex \@plus .2ex}%
	{\normalfont\normalsize\bfseries}}
\newcommand*\l@subsubsubsection{\@dottedtocline{4}{10.0em}{4.1em}}
\newcommand*{\subsubsubsectionmark}[1]{}
\makeatother

\usepackage{ifthen} 
\newboolean{pdflatex}
\setboolean{pdflatex}{true} 

\usepackage{ulem}
\usepackage{caption}
\usepackage{subcaption}
\usepackage{multirow}
\usepackage{amssymb}
\usepackage{longtable}
\usepackage{rotating}
\usepackage{graphicx}
\usepackage{booktabs}
\usepackage{xparse}
\NewDocumentCommand{\rot}{O{45} O{1em} m}{\makebox[#2][l]{\rotatebox{#1}{#3}}}%

\newboolean{articletitles}
\setboolean{articletitles}{true} 

\newboolean{uprightparticles}
\setboolean{uprightparticles}{false} 

\newboolean{inbibliography}
\setboolean{inbibliography}{false} 

\usepackage{longtable} 

\begin{document}

\noindent

{\bf
{\Large 
Analysis of the angular dependence of time delay in gravitational lensing\\

}
} 

\vspace{.5cm}
\hrule

\vspace{1cm}

\noindent

{\large\bf{Nicola Alchera$^{1,2}$, 
Marco Bonici$^{1,2}$, 
Roberta Cardinale$^{1,2}$,
Alba Domi$^{1,2}$,
Nicola Maggiore$^{1,2}$, 
Chiara Righi$^{2,3,4}$,
Silvano Tosi$^{1,2}$  
\\[1cm]}}

\setcounter{footnote}{0}
\hyphenation{obtaining}
\hyphenation{Imposing}
\noindent
{\small{
{{}$^{1}$Dipartimento di Fisica, Universit\`a di Genova, via Dodecaneso 33, I-16146, Ge\-no\-va, Italy\\
{}$^{2}$ I.N.F.N. - Sezione di Genova, via Dodecaneso 33, I-16146, Ge\-no\-va, Italy\\
{$^3$Universit\`a degli studi dell'Insubria, DiSAT, Via Valleggio, 11 - 22100 Como, Italy\\
$^4$INAF - Osservatorio Astronomico di Brera, Via E. Bianchi 46, I-23807 Merate, Italy}
}}
\vspace{1cm}

\noindent
{\tt Abstract~:}

We consider an alternative formula for time delay in gravitational lensing. Imposing a smoothness condition on the gravitationally deformed paths followed by the photons from the source to the observer, we show that our formula displays the same degrees of freedom \textcolor{black}{as} the standard one. In addition to this, it is shown that  the standard expression for time delay is recovered when small angles are involved. These two features strongly support the claim that the formula for time delay studied in this paper is the generalization to \textcolor{black}{the} arbitrary angles of the standard one, which is valid at small angles. This could therefore result in a useful tool \textcolor{black}{in Astrophysics and Cosmology which may be applied to investigate the} discrepancy between the various estimates of the Hubble constant. As an aside, two interesting consequences of our proposal for time delay are discussed: the existence of a constraint on the gravitational potential generated by the lens and a formula for the mass of the lens in the case of central potential.

\newpage

\section{Introduction}

One of the first  modern cosmological models \cite{Einstein:1917ce} was proposed by Albert Einstein in 1917: a static, spatially closed and temporally infinite universe with positive spatial curvature. This model, known as Einstein static universe, although unstable, had the nice feature to be completely determined: the request of staticity fixes the content of the universe, which is composed by matter, curvature and a cosmological constant $\Lambda$. However, in 1929,  Hubble showed that the universe is expanding, ruling out the Einstein model \cite{Hubble:1929ig}. The value of the expansion speed of the universe, later characterized by the Hubble constant $H_0$, was measured to be $500 \mbox{km s}^{-1}\mbox{Mpc}^{-1}$. \textcolor{black}{Then,} the value of $H_0$ was revised along the 20th century, with a controversy about the measurements made by Sandage \cite{1976ApJ...210....7S} (50 $\mbox{km s}^{-1}\mbox{Mpc}^{-1}$) and de Vaucouleurs \cite{1981ApJ...248..395D} (100 $\mbox{km s}^{-1}\mbox{Mpc}^{-1}$). Only in the early 2000s the HST project found a value of $(72\pm8)$ km/s/Mpc \cite{Freedman:2000cf}.

After these works, the measurements of $H_0$ refined more and more, to reach the most recent direct  estimate of the expansion rate of the Universe: $H_0=(73.0\pm1.8) \mbox{km s}^{-1}\mbox{Mpc}^{-1}$ \cite{Riess:2016jrr}. The most important modern techniques arise from the study of the perturbations of the Cosmic Microwave Background (CMB) \cite{Ade:2015xua}, the Supernovae 1A (SN 1A) \cite{Riess:2016jrr} and the effect of Gravitational Lensing (GL) \cite{Suyu:2016qxx,Sluse:2016owq, h0licow3,  Wong:2016dpo,Bonvin:2016crt,  Refsdal:1964nw}. Nevertheless, the different measurements are not compatible one with each other, and still a slight tension about the correct value of $H_0$ does remain \cite{Lukovic:2016ldd}.

The discrepancy could be \textcolor{black}{caused by} a statistical fluctuations or could be evidence of new physics. In order to solve this puzzle, different theoretical scenarios have been proposed. For instance,  it has been showed on general \textcolor{black}{basis}  that dynamical dark energy \cite{DiValentino:2017zyq,Sola:2017znb} or  a specific quintessence model \cite{DiValentino:2017rcr} can solve the tension. Other possibilities have also been considered, such as dark matter-neutrinos interactions \cite{Wilkinson:2014ksa,DiValentino:2017oaw}.

Our contribution towards an attempt to solve the disagreement \cite{Alchera:2017sjt} concerns the theoretical analysis of the time delay $\Delta t$ and its connection to $H_0$. In the standard analysis \cite{falco}, the time delay is calculated adding two contributions: the Shapiro delay from the gravitational potential of the lens and the geometric delay \textcolor{black}{due to} the deformations of the ray paths, approximated by straight lines \cite{falco}. The formula we are considering in this work derives from a different approach which allows to compute the time delay directly in a single shot, rather than two~\cite{Alchera:2017sjt}.

The two formulas for time delay, the standard one \cite{falco} and the one we are proposing \cite{Alchera:2017sjt}, at first sight look quite different. In particular, our proposal seems to depend on more degrees of freedom. In this paper we show that, imposing a reasonable condition of smoothness on the paths of the photons, the parameters of our formula \textcolor{black}{coincide with} those characterizing the standard one, and, most remarkably, we prove that the standard formula is the \textcolor{black}{the small angles limit} of the one we are proposing, \textcolor{black}{so} turns out to be more general.

Furthermore, a weak point of the measurement of $H_0$ through GL is the \textcolor{black}{determination} of the gravitational potential $\Phi$ generated by the lens, which is a crucial quantity in GL, and which, in general \textcolor{black}{is} not a known quantity. \textcolor{black}{This} is, therefore, an important issue \cite{Wong:2016dpo}, which motivated us to study a possible check of consistency for the supposed gravitational potential $\Phi$.

The paper is organized as follows.
In Section \ref{model} we briefly summarize the model underlying our formula for time delay. 
In Section \ref{tangency} we impose a smoothness condition to the paths followed by the photons and we derive the analytical expression of the new parameters in terms of the ones already present in the standard formula. 
The main result achieved in this paper is contained in Section 4, where we show that our formula is the generalization of the standard formula to \textcolor{black}{the} arbitrary angles.
In Section \ref{consistencysec} a consistency check is proposed in order to select a gravitational potential amongst different possibilities.
In Section \ref{mass} we \textcolor{black}{ make the exercise of applying our formula to the simple case of central potential generated by the lens}\textcolor{black}{, obtaining} a formula for the mass of the lens. \textcolor{black}{However}, the mass of the lens \textcolor{black}{is generally}  unknown, being also possibly generated by dark matter. \textcolor{black}{Phenomenological consequences concerning this point, as well as comparison with known results, are also discussed in this Section.}  Our results are summarized in the concluding Section 7.

\section{The model}
\label{model}

In \cite{Alchera:2017sjt} we have obtained a new formula to determine the Hubble constant $H_0$ using time delay $\Delta t$ between multiple images of lensed objects\footnote{Throughout this paper we adopt the notations of \cite{Carroll:2004st}.}
\begin{equation} 
\begin{split}
\Delta t&=\left[ b_2\mu_2-b_1\mu_1\right] +\frac{1}{H_0}\left[(\mathcal{R}(z_{P_2})-\mathcal{R}(z_{Q_2}))     -(\mathcal{R}(z_{P_1})-\mathcal{R}(z_{Q_1})) \right]+\\
&+\frac{1}{H_0}\sum_{k=1}^{+\infty}\left[ \frac{\mathcal{R}(z_{S})\mathcal{R}(z_{Q_2})}{\mathcal{R}(z_{S})-\mathcal{R}(z_{Q_2})}\left( \frac{c_k\gamma_2^{2k}}{2}-\psi_2\right) -\frac{\mathcal{R}(z_{S})\mathcal{R}(z_{Q_1})}{\mathcal{R}(z_{S})-\mathcal{R}(z_{Q_1})}\left( \frac{c_k\gamma_1^{2k}}{2}-\psi_1\right) \right].
\end{split}
\label{our}
\end{equation}
where $b_i$, $P_i$, $Q_i$, $\gamma_i$, $\mu_i$ are parameters defined in Figure \ref{twopaths}, $z_X$ is the redshift of  the generic point $X$ and
\begin{equation}
\mathcal{R}(z_X)\equiv\int_{0}^{z_X}\frac{dz'}{\left[ \sum_{i}\Omega_{i0}(1+z)^{n_i}\right] ^{1/2}},
\label{E}
\end{equation}
where $\Omega_{0i}$ are the four parameters corresponding to radiation, matter, curvature and vacuum and $n_i=4,3,2,0$ respectively. The GL potentials $\psi_i(\vec{\theta})$ ($i=1,2$) are defined as follows
\begin{equation}
\psi_i(\vec{\theta})\equiv2\frac{d_A(LS)}{d_A(EL) d_A(ES)}\int\Phi(d_L\vec{\theta},l) ,
\label{eqn::lensingpot}
\end{equation}
where $d_A(XY)$ is the angular diameter distance of the point $Y$ from the observer $X$. $\Phi$ is the gravitational potential generated by the lens and the integral is \textcolor{black}{performed} over past directed geodesic paths emanating from the observer. Finally, $c_k$ are the coefficients of the Taylor series, which can be easily computed and can be found in \cite{Alchera:2017sjt}.
\par
\begin{figure}[h]
	\includegraphics[scale=0.45]{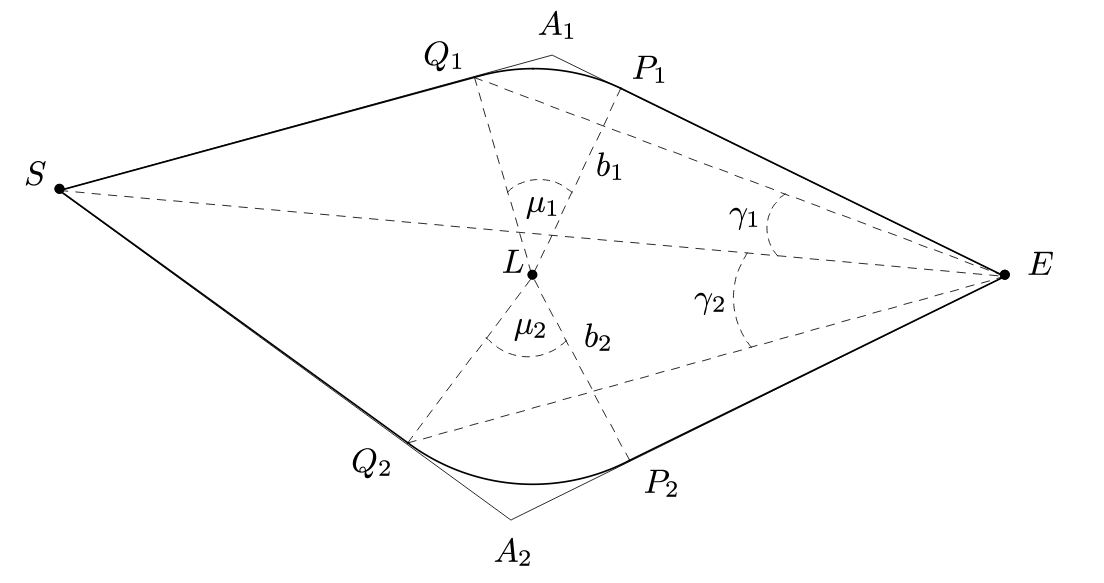}
	\caption{$S$ = source. $L$ = lens. $E$ = observer. $b_i$ = impact parameters. $SQ_iP_iE$ = approximated paths of the photons from $S$ to $E$, indexed by $i=1,2$.}\label{twopaths}
	\centering
\end{figure}
The formula \eqref{our} was built in the framework of a new theoretical model for GL whose geometry is described in Figure \ref{twopaths}. \textcolor{black}{Our approach led to the well known time delay formula \cite{falco}}
\begin{equation}
\Delta t_{old}=\frac{1}{H_0}\frac{\mathcal{R}(z_S)\mathcal{R}(z_L)}{\mathcal{R}(z_S)-\mathcal{R}(z_L)}\left[ \frac{(\alpha_{2}^2-\alpha_{1}^2)}{2}-\left( \psi(\vec{\theta}_2)-\psi(\vec{\theta}_1)\right) \right].
\label{obs}
\end{equation}
\par   
The assumptions on which our model is built are: 
\begin{enumerate}
	\item{The space is divided into two regions: the first, far from $L$, where  $\Phi\approx0$, and the second, close to $L$, where $\Phi \neq 0$.  }
	\item{We approximate the curve $Q_iP_i$ by an arc of a circle centered in $L$. \textcolor{black}{We want to point out that} this is not a necessary condition: the arc of a circle is a good choice to represent $Q_iP_i$, but it is not the only one possible.}
	\item{The universe is spatially flat, \textcolor{black}{compatibly} with observations \cite{Weinberg:2008zzc}.}
\end{enumerate}

According to assumption 1, we chose  $P_i$ and $Q_i$ as the points which divide the photons trajectory $SQ_iP_iE$: from $S$ to $Q_i$ and from $P_i$ to $E$ photons are in the region where $\Phi = 0$, and so they run along straight lines (thanks to assumption 3); from $Q_i$ to $P_i$, instead, photons are in the region where $\Phi \neq 0$, and so they moves on a curved trajectory, which we choose to describe with the arc of a circle $Q_iP_i$ in Figure \ref{twopaths}.

There are at least  two different geodesics along which photons can move from $S$ to $E$ and this causes the time delay $\Delta t$ \cite{falco}.\par
These assumptions led us to the time delay formula \eqref{obs}, which relates $\Delta t$ to $H_0$.

\section{Smoothness condition}\label{tangency}

In \cite{Alchera:2017sjt} we emphasized that the formula \eqref{our}, as it stands,  is not well suited for actual calculations of time delay, in the hope of softening the discrepancy among the existing estimates of $H_0$. The reason is that the geometry is not uniquely defined because we have placed no constraints on $Q_i$, $P_i$, $\mu_i$, $b_i$ and $\gamma_i$ which, consequently, are free parameters. 
\textcolor{black}{It is easy to show that} imposing \textcolor{black}{on} a smoothness condition of the paths $SQ_iP_iE$ uniquely fixes the parameters appearing in the new formula \eqref{our} which, therefore, are not free. The next task is to show that \eqref{our}} can be written in terms of the parameters appearing in the usual formula \eqref{obs}. In other terms, no new degrees of freedom have been introduced in our reformulation of the time delay formula. Moreover, expressing the new formula by means of the parameters appearing in \eqref{obs}, makes easier the comparison of the two results, rendering\textcolor{black}{,} at the same time\textcolor{black}{,} more clear the domain of application of our new approach.
Let us consider for the moment only one path, as in Figure \ref{geometria}. The equation of the straight line passing through $EP$ is
\begin{equation} \label{rettaEP}
y=x\tan \theta .
\end{equation}
The equation of the circle centered in $L$ of radius $b$ to which belongs the arc $QP$ is
\begin{equation}  \label{circumference}
(x-x_L)^2+y^2=b^2.
\end{equation}
From $\hat {LPE}= \pi/2$  we have
\begin{equation}\label{impact}
b=x_L \sin \theta,
\end{equation}
so that the coordinates of $P$ are
\begin{gather}\label{pointP}
x_P= EP \cos \theta=x_L \cos^2 \theta, \\
y_P=EP \sin \theta= x_L \cos\theta \sin \theta, 
\end{gather}
where we have used  $EP= x_L \cos \theta$ and the relation \eqref{impact}. 

\begin{figure}[h]
	\includegraphics[scale=0.4]{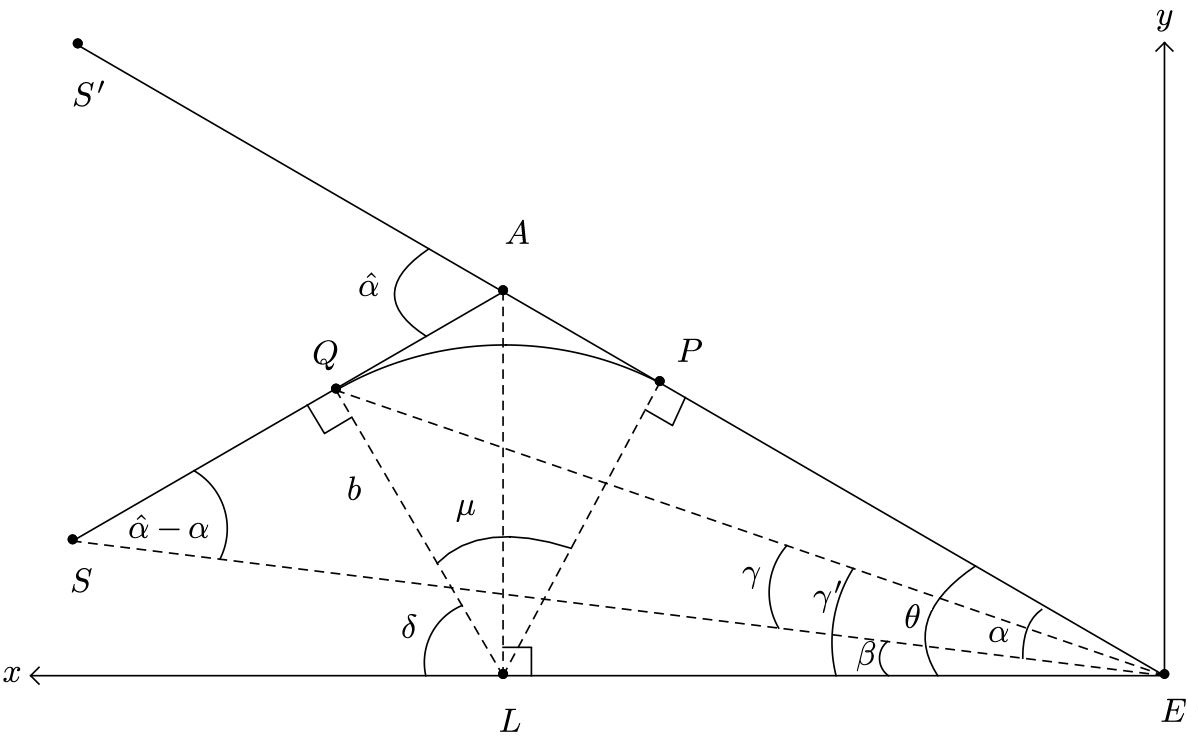}
	\centering
\caption{\textcolor{black}{The geometry within which we will develop our calculation. Only one path is displayed.}}
\label{geometria}
\end{figure}
Analogously, the equation of the straight line  $QL$ is
\begin{equation} \label{LQ}
y= (x-x_L)\tan \delta,
\end{equation}
where $\delta$  is defined as in Figure \ref{geometria}, \textcolor{black}{which makes clear} that

\begin{equation}
\delta=\frac{\pi}{2}+\theta-\hat \alpha.
\label{delta}
\end{equation}
Intersecting the line \eqref{LQ} with the circle \eqref{circumference}  we obtain
\begin{equation}
(x-x_L)^2=x_L^2\sin^2\theta\cos^2\delta,
\label{}\end{equation}
\textcolor{black}{and consequently}
\begin{equation}
x_Q=\left( 1+\cos\delta\sin\theta\right)  x_L\qquad y_Q=\sin\delta\sin\theta x_L.
\label{Q}
\end{equation}

Now we are able to obtain the value of $\gamma'$ as defined in Figure  \ref{geometria}
\begin{equation}
\gamma'=\arctan\left( \frac{\sin\theta\sin\delta}{1+\cos\delta\sin\theta}\right) .
\end{equation}
Now we observe that
\begin{equation}
\gamma'=\gamma+\alpha-\theta,
\end{equation}
hence
\begin{equation}
\gamma=\arctan\left( \frac{\sin\theta\sin\delta}{1+\cos\delta\sin\theta}\right)+\alpha-\theta,
\label{gamma}
\end{equation}
and the angles $\alpha$ and $\hat{\alpha}$ are related by \cite{Carroll:2004st}
\begin{equation}
\alpha=\frac{d_A(LS)}{d_A(ES)}\widehat\alpha.
\end{equation}
We have now all the tools to write the formula \eqref{our} in terms of the parameters appearing in the usual one \eqref{obs}.
Let us rewrite \eqref{our} as 
\begin{equation}\label{terms}
\Delta t= \Delta\tau_1 + \Delta\tau_2 + \Delta\tau_3,
\end{equation}
where
\begin{equation}
	\Delta\tau_1\equiv\left[ b_2\mu_2-b_1\mu_1\right],
\end{equation}
\begin{equation}
	 \Delta\tau_2 \equiv\frac{1}{H_0}\left[(\mathcal{R}(z_{P_2})-\mathcal{R}(z_{Q_2}))     -(\mathcal{R}(z_{P_1})-\mathcal{R}(z_{Q_1})) \right],
	 \label{A2def}
\end{equation}
\begin{equation}
	\Delta\tau_3\equiv\frac{1}{H_0}\sum_{k=1}^{+\infty}\left[ \frac{\mathcal{R}(z_{S})\mathcal{R}(z_{Q_2})}{\mathcal{R}(z_{S})-\mathcal{R}(z_{Q_2})}\left( \frac{c_k\gamma_2^{2k}}{2}-\psi_2\right) -\frac{\mathcal{R}(z_{S})\mathcal{R}(z_{Q_1})}{\mathcal{R}(z_{S})-\mathcal{R}(z_{Q_1})}\left( \frac{c_k\gamma_1^{2k}}{2}-\psi_1\right) \right].
	\label{A3def}
\end{equation}
Using  \eqref{impact} we can write the first contribution to $\Delta t$ as
\begin{equation}
\Delta\tau_1=x_L(\sin\theta_2\hat{\alpha}_2-\sin\theta_1\hat{\alpha}_1).
\label{A1}
\end{equation}
Let us now focus on the $\Delta\tau_2$ term.
From \cite{Alchera:2017sjt}
\begin{equation}
\mathcal{R}(z_{X})=H_0r_{X},
\end{equation}
and \eqref{Q}, we have
\begin{eqnarray} 
\mathcal{R}(z_Q)&=&\mathcal{R}(z_L)\sqrt{1+\sin^2\theta-2\sin(\theta-\hat{\alpha})\sin\theta} \label{RQ} 
\\
\mathcal{R}(z_{P}) &=& H_0r_{P} = H_0 \cos\theta x_L=\mathcal{R}(z_L)\cos\theta \label{RP}. 
\end{eqnarray}
We are \textcolor{black}{then} able to write $\Delta\tau_2$ in \eqref{terms} as follows
\begin{eqnarray}
\Delta\tau_2 
&=&
\frac{1}{H_0}\mathcal{R}(z_{L})  
(\cos\theta_2-\sqrt{1+\sin^2\theta_2-2\sin(\theta_2-\hat{\alpha}_2)\sin\theta_2} 
\nonumber \\
&&-\cos\theta_1+\sqrt{1+\sin^2\theta_1-2\sin(\theta_1-\hat{\alpha}_1)\sin\theta_1}).
\label{A2}
\end{eqnarray}
Finally, using \eqref{gamma} and \eqref{Q}, we have
\begin{equation}
\begin{split}
\Delta\tau_3=&\frac{1}{H_0}\sum_{k=1}^{+\infty}(\frac{\mathcal{R}(z_S)\mathcal{R}(z_L)\sqrt{1+\sin^2\theta_2-2\sin(\theta_2-\hat{\alpha}_2)\sin\theta_2}}{\mathcal{R}(z_S)+\mathcal{R}(z_L)\sqrt{1+\sin^2\theta_2-2\sin(\theta_2-\hat{\alpha}_2)\sin\theta_2}} \\ &\frac{c_k}{2}\left( \arctan\left( \frac{\sin\theta_2\cos(\hat{\alpha_2}-\theta_2)}{1-\sin(\theta_2-\hat{\alpha}_2)\sin\theta_2}\right)+\alpha_2-\theta_2\right)^2+\\
&\frac{\mathcal{R}(z_S)\mathcal{R}(z_L)\sqrt{1+\sin^2\theta_1-2\sin(\theta_1-\hat{\alpha}_1)\sin\theta_1}}{\mathcal{R}(z_S)+\mathcal{R}(z_L)\sqrt{1+\sin^2\theta_1-2\sin(\theta_1-\hat{\alpha}_1)\sin\theta_1}} \\ &\frac{c_k}{2}\left( \arctan\left( \frac{\sin\theta_1\cos(\hat{\alpha}_1-\theta_1)}{1-\sin(\theta_1-\hat{\alpha}_1)\sin\theta_1}\right)+\alpha_1-\theta_1\right)^2).
\end{split}\label{A3}
\end{equation}

The exact expression for the time delay \eqref{our}, using \eqref{terms}, is then given by the sum of the three terms $\Delta\tau_i$  \eqref{A1}, \eqref{A2} and \eqref{A3}. We stress that the two expressions for time delay $\Delta t$ \eqref{our} and $\Delta t_{old}$ \eqref{obs} are expressed by means of the same parameters: no new degrees of freedom have been introduced.

\textcolor{black}{
We conclude this Section with a few words concerning the possibility of a comparison with other proposals of alternative formulae for the time delay. The most important quests where gravitational time delay occurs are the determination of the Hubble constant \cite{Bonvin:2016crt} and the tests of General Relativity \cite{Shapiro:1964uw}. In all these cases the formula which is used is the one appearing in \cite{falco}, namely Eq. (2.4). At most, the formula (2.4) is heuristically modified, as done in \cite{Bonvin:2016crt} in order to take into account the multiple galaxies at different redshifts close in projection to the strong lens system. Actually, formulae for time delay have also been given for very particular cases, like, for instance, in \cite{Virbhadra:2007kw} for lensing by Schwarzschild black hole and naked singularities, and in \cite{Jusufi:2017drg} where is considered the time delay generated by black holes and massless wormholes in massive gravity, but the interest of these latter expressions for time delay is mainly formal, rather than phenomenological. In this paper, instead, we propose the first attempt to go beyond the standard formula for time delay, resting on basic grounds of General Relativity. As a matter of fact, we rely only on the definitions of angular distance and of redshift in Cosmology. Therefore, for what concerns the formula presented in this paper, the most pertinent comparison is only with the standard formula (2.4).
}

\section{Small angles limit }\label{convergence}

In this Section we \textcolor{black}{are showing} that  the standard formula for time delay \eqref{obs}  is recovered in the small angles limit of our formula \eqref{our}. This is a remarkable result, because it proves that the formula \eqref{our} is the generalization of the standard \textcolor{black}{one} to \textcolor{black}{the} arbitrary angles. In order to get this result, the crucial step is that the formula $\Delta t$, written in the form \eqref{terms}, is expressed in terms of the same degrees of freedom appearing in the standard formula \eqref{obs}.

Let us consider the equation \eqref{gamma} and let us suppose that all the angles involved are very small. Up to the second non vanishing order, we get
\begin{equation}
\gamma\simeq \alpha-\frac{1}{2}\theta\widehat\alpha^2.
\label{gammasecond}
\end{equation}
\textcolor{black}{Now, we} expand $\mathcal {R}(z_{Q})$ and $\mathcal {R}(z_{P})$, respectively given by \eqref{RQ} and \eqref{RP}. We obtain
 
\begin{eqnarray}
\mathcal{R}(z_{Q})&=&H_0r_{Q}\simeq H_0(1+\hat{\alpha}\theta-\frac{\theta^2}{2})x_L,
\label{limRq}\\
\mathcal{R}(z_{P})&=&H_0r_{P}\simeq H_0\left ( 1-\frac{\theta^2}{2}\right) x_L.
\label{limRp}
\end{eqnarray}
Now we are able to expand  in the small angles limit the time delay formula \eqref{our} term by term, according to \eqref{terms}.
For what concerns $\Delta\tau_1$ \eqref{A1}, we have
\begin{equation}\label{A1approx}
\Delta\tau_1 \simeq x_L(\hat{\alpha}_2\theta_2-\hat{\alpha}_1\theta_1).
\end{equation}
Let us consider the $\Delta\tau_2$ term given by \eqref{A2def}. Using \eqref{limRq} and \eqref{limRp} we obtain
\begin{equation}\label{A2approx}
\begin{split}
\Delta\tau_2\simeq -x_L(\hat{\alpha}_2\theta_2-\hat{\alpha}_1\theta_1).
\end{split}
\end{equation}
Finally, using \eqref{gammasecond} and \eqref{limRq} we obtain
\begin{equation}\label{approx}
\frac{\mathcal{R}(z_{S})\mathcal{R}(z_{Q_i})}{\mathcal{R}(z_{S})+\mathcal{R}(z_{Q_i})}\gamma_i^2
\simeq
\frac{\mathcal{R}(z_{S})\mathcal{R}(z_{L})}{\mathcal{R}(z_{S})+\mathcal{R}(z_{L})}\alpha_i^2.
\end{equation}
Plugging \eqref{approx} and \eqref{gammasecond} in \eqref{A3def} we obtain for the last contribution $\Delta\tau_3$
\begin{equation}\label{A3approx}
\Delta\tau_3\simeq\frac{1}{H_0}\frac{\mathcal{R}(z_S)\mathcal{R}(z_L)}{\mathcal{R}(z_S)-\mathcal{R}(z_L)}\left[ \frac{(\alpha_{2}^2-\alpha_{1}^2)}{2}-\left( \psi(\vec{\theta}_2)-\psi(\vec{\theta}_1)\right) \right].
\end{equation}
Using \eqref{A1approx}, \eqref{A2approx} and \eqref{A3approx} in the equation \eqref{terms} we obtain the following formula 
\begin{equation}
\begin{split}
\Delta t = \frac{1}{H_0}\frac{\mathcal{R}(z_S)\mathcal{R}(z_L)}{\mathcal{R}(z_S)-\mathcal{R}(z_L)}\left[ \frac{(\alpha_{2}^2-\alpha_{1}^2)}{2}-\left( \psi(\vec{\theta}_2)-\psi(\vec{\theta}_1)\right) \right]+ O(\theta^3),
\end{split}
\label{abo4}
\end{equation}
which is precisely the formula already present in literature \eqref{obs} and currently used in the determination of $H_0$ through GL. We can therefore conclude that

\begin{equation}
\lim_{\alpha,\theta\rightarrow 0}\Delta t = \Delta t_{old} + O(\theta^3).
\end{equation}
This remarkable fact justifies our claim: our time delay formula \eqref{terms}, which is equivalent to \eqref{our} once the smoothness condition is imposed, is the generalization \textcolor{black}{at the} arbitrary angles of the standard expression \eqref{obs}, and, differently from \eqref{obs}, it has been obtained in one shot only without \textcolor{black}{having to consider} two distinct steps (geometrical and Shapiro).

\textcolor{black}{
Some comments are in order concerning the relation between the result presented in [20] and the one obtained in this paper.
In [20], the formula (2.1) for time delay has been proposed which, a priori, was supposed to be an alternative to the standard one (2.4). It was indeed derived in a different way, in particular without adding together two independent contributions, namely the Shapiro and the geometrical ones, but, rather, by means of a single reasoning, namely a refinement of the approximation concerning the path of the photon in its way from the source through the observer. The resulting formula (2.1) appears to depend on more degrees of freedom than the standard one, and only for a particular, arbitrary albeit reasonable, choice of the parameters, the two formulae were shown to coincide. This, evidently, is not sufficient to claim that the new formula generalizes the standard one. At most, it is possible to conclude that the two formulae coincide in a subset of the space of parameters. In this paper we do not make any assumption on the parameters. Instead, we impose a condition of smoothness on the path followed by the photons. Doing this, quite unexpectedly, we find the remarkable result that the degrees of freedom collapse into the standard ones. This is the first clue that the new formula is a generalization of the old one, rather than something different and alternative. The definitive proof of this statement is contained in Section 4, where it is shown that, going to small angles, the standard formula is recovered. This demonstration is the main achievement of this paper. Of course, the breakthrough would be the use of this more general formula in physical situations where the angles involved are not small, and therefore the standard formula cannot be used. At the moment, however, the actual technology does not allow the quantitative observation of the phenomenon of gravitational lensing at large, or at least not too small, angles.
}

\section{A constraint on the lens gravitational potential} \label{consistencysec}

The time delay formula \eqref{obs} or, equivalently, \eqref{our}, allows to determine $H_0$. The crucial physical quantity for the time delay is the gravitational potential $\Phi$ originated by the lens $L$. In most cases, $\Phi$ is not known but rather is assumed to belong to a certain class, or inferred by some theoretical considerations \cite{Wong:2016dpo}. We will denote with $\Phi_{hp}$ the assumed potential, which, hopefully, should not be very far from the real one, which we shall call $\Phi_{phys}$. Our aim is to provide a consistency relation for $\Phi_{hp}$, by means of which it will be possible to check \textcolor{black}{if} $\Phi_{hp}$ represents a realistic assumption for the real gravitational potential $\Phi_{phys}$ of the lens or not. From now on we will use the subscript ``\textit{hp}'' for the quantities deduced from $\Phi_{hp}$ and the subscript ``\textit{phys}'' for their actual values.
Once that the potential $\Phi_{hp}$ has been assumed, from \cite{Carroll:2004st}
\begin{equation}
	\hat{\alpha}=2\int\nabla_\perp\Phi ds,
\end{equation}
where $\nabla_\perp\Phi$ is the transverse gradient of the potential with respect to the path, we can compute the angle $\hat{\alpha}_{hp}$ which, consequently, not necessarily coincides with $\hat{\alpha}_{phys}$. From the experimental knowledge of $\theta_i$ and of $z_L$, and once that $(\hat{\alpha}_i)_{hp}$ are determined, we can predict the position of the source $(\vec{r}_S)_{hp}$. An obvious check for $\Phi_{hp}$ would be
\begin{equation}
	(\vec{r}_S)_{hp}=(\vec{r}_S)_{phys},
\end{equation}
where $(\vec{r}_S)_{phys}$ identifies the actual position of the source $S$.
Unfortunately, the exact position of the source $(\vec{r}_S)_{phys}$ \textcolor{black}{is rarely} known. What \textcolor{black}{is generally known}, \textcolor{black}{instead}, is its redshift $z_S$, from which, using \eqref{E}, the distance $\mid(\vec{r}_S)_{phys}\mid\equiv (r_S)_{phys}$ can be computed. Therefore, the check of consistency reduces to
\begin{equation}
	(r_S)_{hp}=(r_S)_{phys},
	\label{constraint}
\end{equation}
which is a necessary condition for $\Phi_{hp}$. We can compute $(r_S)_{phys}$ explicitly as follows.\par
From \eqref{Q} and \eqref{delta} we have
\begin{equation}
x_{Q_i}=\left( 1+\sin((\hat{\alpha}_i)_{hp}-\theta_i)\sin\theta_i\right)x_L \qquad y_{Q_i}=\sin\theta_i\cos((\hat{\alpha}_i)_{hp}-\theta_i)x_L,
\end{equation}
and the equations of the straight lines passing through $Q_i$ and $S$ are
\begin{equation}
y-y_{Q_i}=-\tan((\hat{\alpha}_i)_{hp}-\theta_i)(x-x_{Q_i}),
\label{Q1S}
\end{equation}
where, again, $i=1,2$.
From \eqref{Q1S} we get the coordinates of the source $S$
\begin{equation}
(x_S)_{hp}=\frac{y_{Q_1}-y_{Q_2}+\tan((\hat{\alpha}_1)_{hp}-\theta_1)x_{Q_1}+\tan((\hat{\alpha}_2)_{hp}-\theta_2)x_{Q_2}    }{\tan((\hat{\alpha}_1)_{hp}-\theta_1)+\tan((\hat{\alpha}_2)_{hp}-\theta_2)},
\end{equation}
\begin{equation}
\begin{split}
(y_S)_{hp}&=y_{Q_1}+\tan((\hat{\alpha}_1)_{hp}-\theta_1)x_L-\tan((\hat{\alpha}_1)_{hp}-\theta_1)x_S.
\end{split}
\end{equation}
The \textcolor{black}{small angles limit} $\hat{\alpha}_i$ and $\theta_i$ gives
\begin{equation}
(x_S)_{hp}=\frac{(\hat{\alpha_1})_{hp}+(\hat{\alpha_2})_{hp}}{(\hat{\alpha_1})_{hp}+(\hat{\alpha_2})_{hp}-\theta_1-\theta_2}x_L,
\label{XS}
\end{equation}
\begin{equation}
(y_S)_{hp}=\frac{(\hat{\alpha_2})_{hp}\theta_1-(\hat{\alpha_1})_{hp}\theta_2}{(\hat{\alpha_1})_{hp}+(\hat{\alpha_2})_{hp}-\theta_1-\theta_2}x_L,
\label{YS}
\end{equation}
and hence the distance of the source $(r_S)_{hp}$ is
\begin{align}
(r_S)_{hp}=\frac{\sqrt{(\hat{\alpha}_1)_{hp}^2(1+\theta_2^2)+(\hat{\alpha}_2)_{hp}^2(1+\theta_1^2)+2(\hat{\alpha}_1)_{hp}((\hat{\alpha_2})_{hp})_{hp}(1-\theta_1\theta_2)}}{(\hat{\alpha_1})_{hp}+(\hat{\alpha_2})_{hp}-\theta_1-\theta_2}x_L.
\label{rs2}
\end{align}
This quantity depends on the assumed choice of the gravitational potential $\Phi_{hp}$ generated by the lens $L$ through the angles $(\hat{\alpha}_i)_{hp}$. Therefore, once that the expression \eqref{rs2} for $(r_S)_{hp}$ is given, the validity of the constraint \eqref{constraint} can be checked.

\section{Determination of the lens mass for central potential} \label{mass}

\textcolor{black}{In order to give a simple example of how our \textcolor{black}{time delay} formula can be used for phenomenological calculations, let us consider the textbook case of spherically symmetric potential $\Phi$, being aware that this is not a realistic assumption for actual gravitational potentials generated by the lens.} It is a known result  \cite{Carroll:2004st} that in this case the angles are given by
\begin{equation}
\hat{\alpha}_i=\frac{4GM}{d_{EL}\theta_i},
\label{sphere}
\end{equation}
where $M$ is the mass of the lens $L$ and $d_{EL}$ is the angular diameter distance between $L$ and $E$. Inserting  \eqref{sphere} in \eqref{rs2} we obtain
\begin{equation}
\begin{split}\label{distancemass}
r_S=\frac{4GM\sqrt{(\theta_1+\theta_2)^2+(\theta_1^2-\theta^2_2)^2}}{(4GM-d_{EL}\theta_1\theta_2)(\theta_1+\theta_2)}x_L,
\end{split}
\end{equation}
and hence
\begin{equation}
M=\frac{\theta_1\theta_2d_{EL}r_S}{4G\left( r_{ES}-\left( 1+\frac{(\theta_1-\theta_2)^2}{2} \right) x_L\right) }.
\end{equation}
Using the relation between the angular diameter distance $d$ and the proper distance $r$ \cite{Weinberg:2008zzc}
\begin{equation}
d_{EL}=a(t_L)r_{EL},
\end{equation}
we can conclude that
\begin{align}\label{massasferica}
M=\frac{\theta_1\theta_2d_{EL}d_{ES}}{4G\left( d_{ES}-\frac{1+z_L}{1+z_S}\left(1+\frac{(\theta_1-\theta_2)^2}{2}\right)d_{EL}\right) }.
\end{align}

This is a general result, valid for all spherically symmetric lenses. \textcolor{black}{The lesson from this simple example, is that in principle} it is possible to estimate the mass of a lens with central gravitational potential, like a star or a black hole, once the angular diameter distances $d_{ES}$, $d_{EL}$, the observed angle $\hat \theta_i$ and the redshifts $z_L,\ z_S$ are known. \textcolor{black}{It is interesting to compare the results coming from the mass formula  \eqref{massasferica} with those obtained using the corresponding formula existing in literature.}

If a circularly symmetric lens is considered and if lens, source and observer arecollinear, as a consequence of the rotational symmetry of the lens system\textcolor{black}{,} the source is imaged as a ring. The radius of the ring, called Einstein radius, is
given by
\begin{equation}
\theta_{E} = [{4 G M(\theta_{E})} \,
\frac{d_{LS}}{d_{EL}d_{ES}}]^{1/2}
\label{eq:thetaEinstein}
\end{equation}
where $d_{EL}$, $d_{ES}$ and $d_{LS}$ are the angular diameter
distances. 
From \eqref{eq:thetaEinstein}, it is possible to estimate the mass
inside the Einstein ring, which is given by
\begin{equation}
M(\theta_E) = \frac{1}{4 G } \,
\frac{d_{EL}d_{ES}}{d_{LS}} \theta_{E}^{2}.
\label{eq:massEinstein}
\end{equation}
This is the standard formula provided in
literature (see for example~\cite{schmidt}).
If the source and the lens are misaligned, multiple images can be
observed. 
In this case it is possible to measure the relative distances between each image and the lens, called $\theta_{1}$ and $\theta_{2}$ for a two-images lens system, but it is also possible to extract an ``effective'' Einstein radius by performing a fit assuming a particular mass distribution of the lens.
It has been shown that the assumed mass distribution model only partly affects
the Einstein ring extraction~\cite{Oguri:2013mxl,
	Suyu:2012rh}, and since the Einstein
angle is an average between the lens center and multiple
images, in the case of a two images system the $\theta_{E}$ angle is usually replaced with $\theta_{E} =
\frac{(\theta_{1}+\theta_{2})}{2}$.
\newline
This is usually a good approximation for systems with nearly
symmetric image configuration ($\theta_{1}/\theta_{2} \lessapprox 2$)
where the mass enclosed within $\theta_{E}$ is accurate to within

$\sim 5\%$ as reported in~\cite{Suyu:2012rh}. 
The equation \eqref{massasferica} is indeed a generalisation of the standard formula and reduces to the standard formula if $\theta_{1} = \theta_{2}$. 

In order to test the validity of the lens mass estimation, a comparison between the standard formula \eqref{eq:thetaEinstein} and the formula for the mass of the lens obtained in this paper \eqref{massasferica}, which uses explicitly $\theta_{1}$ and $\theta_{2}$, is provided.
To have a fair comparison and to verify the applicability of \eqref{massasferica}, the formula is applied to a subset of quasar lenses listed in the CASTLES webpage~\cite{castles} satisfying the following requirements: first of all, the source and the lens redshifts have to be known. Complex lens systems such as multiple galaxies or cluster of galaxies are excluded and only quasar lenses with a planar geometry with two
lens images are chosen. Finally, the observed angles $\theta_{1}$ and  $\theta_{2}$ must be known data.
Five quasar lenses have been identified to fulfill the previous
criteria. 
Our study shows that in the case, $\theta_{1}/\theta_{2} \lessapprox 2$, the
standard formula is a good approximation of the formula \eqref{massasferica} and as it
can be seen in Table~\ref{tab:masscomparison}: for
QJ0158-4325 the correction is negligible, while for SDSS1226-0006  the correction is of the order of $10\%$. 

For $\theta_{1}/{\theta_{2}} > 2$, the standard formula starts to
show a significant discrepancy with respect to the formula obtained in
this paper. The limit of the standard formula is overcome by equation
\eqref{massasferica} that can indeed be applied for any angle configuration. 

An ideal outcome of our study would be the determination of the dark matter fraction in the lens galaxy in a more precise way. To do that, a more appropriate potential has of course to be used to take into account the density distribution of dark matter in a galaxy. \textcolor{black}{This} additional study represents a future development of our work.


The parameters for the five considered lenses are reported in
Table~\ref{tab:lensparam}.
\begin{table}[htbp]
	\centering
	\begin{tabular}{l|r|r|r|r|r}
		Lens & $z_{L}$ & $z_{S}$ & $\theta_{1}$ & $\theta_{2}$ &$\theta_{E}$\\
		&           &            &  [$10^{-6}$ rad] & [$10^{-6}$ rad] & [$10^{-6}$ rad]\\
		\hline
		QJ0158-4325 & 0.317 & 1.29    & $3.95 \pm 0.07$          & $1.99 \pm 0.07$      &  $2.8$ \\
		J1004+1229 & 0.95 & 2.65        & $6.156 \pm 0.034$     & $1.309 \pm 0.037$ & $4.02$\\
		HE1104-1805 & 0.73 & 2.32      & $10.216 \pm 0.021$   & $5.269 \pm 0.015$ & $6.8$\\
		SDSS1226-0006 & 0.52 & 1.12 & $3.992 \pm 0.021$     & $2.120 \pm 0.021$ & $2.76$\\
		HE2149-2745 & 0.5 & 2.03        & $6.563 \pm 0.027$      & $1.670 \pm 0.031$ & $4.1$\\
		\hline
	\end{tabular}
	\caption{}
	\label{tab:lensparam}
\end{table}
We considered a flat $\Lambda$CDM universe, with $\Omega_M=0.3$, $\Omega_\Lambda=0.7$ and $H_0=70$ km/s/Mpc.
The calculated masses using the parameters reported in
Table~\ref{tab:lensparam} are reported in
Table~\ref{tab:masscomparison}, \textcolor{black}{where $M_{\text{our}}$ and $M_{\text{std}}$ are calculated using \eqref{massasferica}  and \eqref{eq:massEinstein}, respectively}.
\begin{table}[htbp]
	\centering
	\begin{tabular}{c|c|c|c}
		Lens & $M_{\text{std}}$ & $M_{\text{our}}$ & $100\cdot(M_{\text{std}}-M_{\text{our}})/M_{\text{std}}$\\
		&   [$M_\odot$] & [$M_\odot$] & \\
		\hline
			QJ0158-4325  &$5.80\cdot10^{10}$&$5.81\cdot10^{10}$&$-0.17$\\
		J1004+1229&$2.97\cdot10^{11}$&$1.48\cdot10^{11}$&50.13\\
		HE1104-1805 &$6.81\cdot10^{11}$&$7.93\cdot10^{11}$&$-16.41$\\
		SDSS1226-0006& $1.14\cdot10^{11}$&$1.26\cdot10^{11}$&$-11.11$\\
		HE2149-2745 &$1.75\cdot10^{11}$&$1.14\cdot10^{11}$&34.82\\
		\hline
	\end{tabular}
	\caption{}
	\label{tab:masscomparison}
\end{table} 

\textcolor{black}{We stress that
the results contained in Sections 5 and 6, are secondary with respect to the main one obtained in the previous Sections, namely the generalized formula for time delay. These are mathematical achievements with predictive consequences, and, in our opinion, represent interesting applications of the new formula for time delay, with possible phenomenological implications. The first is a no-go theorem for the possible gravitational potentials generated by the lens, whose determination is currently object of intensive research. The second consequence is the determination of Eq. (6.5) for the total mass which deviates the path of the photon. By total we mean the sum of bot visible and dark matter. This formula, which holds for the particular case of spherical symmetric mass distributions, has been compared with Eq. (6.7), which is the one currently used in the same situations, of which the new mass formula appears to be a refinement. In the few cases we considered as a test, we found interesting deviations from the existing estimates, which deserve further investigations.
}

\section{Conclusions}

In this paper we improved the analysis of the time delay contained in a previous work \cite{Alchera:2017sjt}, where we obtained a new expression for $\Delta t$, characterized, however, by the presence of a number of apparently free parameters, whose presence renders that expression unsuitable for phenomenological considerations. \textcolor{black}{Altough}, as we showed, its validity is enforced by the fact that, for certain values of its parameters, it reduces to the standard one \cite{falco}. Both in the standard approach and in ours, the paths followed by the photons from the source to the observer, modified by the presence of the lens, were approximated either by straight lines (as in \cite{falco}) or by straight lines and an arc of a circle (as in \cite{Alchera:2017sjt}), not joining in a smooth way.\\ \textcolor{black}{Two are the main results presented in this paper}. \textcolor{black}{The first is that}, imposing a smoothness condition on the photon rays, the number of free parameters drastically reduces, with the outcome that no new degrees of freedom are introduced with respect to the standard formula for time delay. \textcolor{black}{The second outcome is represented by the fact that} we were able to show that our formula exactly reduces to the standard one in the \textcolor{black}{small angles limit}, which means that our expression for time delay generalizes the standard one to generic angles. In addition, we gave a consistency check for the gravitational potential generated by the lens. Although this physical quantity is crucial for the GL effect, and in particular for the determination of the Hubble constant $H_0$ by means of the time delay formula, its exact expression is rarely known, and all the considerations are made on the basis of conjectured potentials. For instance, the GL effect could be generated also by dark matter distributions, of which, at most, only hypothetical maps exist. A criterion for selecting amongst different guesses on the gravitational potential generating the GL effect is therefore useful. \\
Finally, \textcolor{black}{to give a taste of the possible applications of our formula}, we applied our results to the case of a central gravitational potential, obtaining an expression for the mass of the lens, which, again, \textcolor{black}{is generally} unknown. \textcolor{black}{Although different situations exist, for which the spherical symmetry is a  good approximation, realistic gravitational potentials generated by the lenses are  in general far more complicated. Our aim here is only to show which might be the phenomenological applications of our formula. 
The signal of the particular case of  a spherically symmetric gravitational potential is the presence of only two images, which are aligned with the lens.}
\textcolor{black}{We analyzed five situations where our formula \eqref{massasferica} for the mass can be successfully applied, and we compared our results to those obtained with the mass formula traditionally used in case of central potentials \eqref{eq:massEinstein}. The promising outcome is that our formula reproduces the results obtained in the usual way, but, quite remarkably, also extends the range of application of the standard one to any angle configuration.}
This approach could give hints on the dark matter presence in high $z$ galaxies: from their luminosity it is possible to estimate the value of their mass, which could be compared to our theoretical prediction. This would be interesting in situations where it is not possible to use standard methods, such as the study of the galaxy rotations curve, due to the high distances involved.

\par
{\bf Acknowledgements}

N.M. thanks the support of INFN Scientific Initiative SFT: ``Statistical Field Theory, Low-Dimensional Systems, Integrable Models and Applications''. Luca Panizzi and Davide Ricci are gratefully acknowledged. We thank Laura Massa Trucat for a careful reading of the manuscript.


\end{document}